\documentclass[10pts,showpacs,preprint,aps]{revtex4}
\usepackage{graphicx}% Include figure files
\usepackage{dcolumn}% Align table columns on decimal point
\usepackage{bm}% bold math
\usepackage{amssymb}
\usepackage{amsmath}
\begin{document}
\setcounter{page}{1}
\vskip 2cm
\title
{Spontaneous symmetry breaking as a triangular relation between pairs of Goldstone bosons and the degenerate vacuum: Interactions of D-branes}
\author
{Ivan Arraut$^{(1)}$}
\affiliation{$^1$Department of Physics, Faculty of Science, Tokyo University of Science,
1-3, Kagurazaka, Shinjuku-ku, Tokyo 162-8601, Japan}

\begin{abstract}
We formulate the Nambu-Goldstone theorem as a triangular relation between pairs of Goldstone bosons with the degenerate vacuum. The vacuum degeneracy is then a natural consequence of this relation. Inside the scenario of String Theory, we then find that there is a correspondence between the way how the $D$-branes interact and the properties of the Goldstone bosons.                         
\end{abstract}
\maketitle 
\section{Introduction}
For a system having two or more broken generators satisfying a Lie algebra, it is possible to formulate the principle of spontaneous symmetry breaking as a triangular relation between pairs of Goldstone bosons and the degenerate vacuum \cite{YB}. From the triangular relation, the vacuum degeneracy emerges naturally. In this formulation, the Goldstone bosons take a fundamental role over the broken generators. The standard formulation of the Nambu-Goldstone theorem suggests that the number of Goldstone bosons is equal to the number of broken generators ($N_{NG}=N_{BG}$) \cite{Goldstone}. Then whether we consider as a fundamental variable the broken symmetries of the system or the Goldstone bosons associated to them is a matter of choice. For internal symmetries, when the broken generators form conjugate pairs, it has been demonstrated that some correction is necessary in order to find the appropriate number of Goldstone bosons \cite{YB, Murayama, MOP}. In addition, the dispersion relations for the Goldstone bosons coming from conjugate pairs of broken generators is not linear as it should be expected from massless particles. This however does not mean that the Nambu-Goldstone theorem fails. This means that under some circumstances pairs of Goldstone bosons become effectively a single degree of freedom with quadratic dispersion relation. This result can only be obtained naturally from the triangular relation proposed in \cite{YB}. In this paper we explain in detail the results obtained in \cite{YB} by starting from the fundamental interaction of pairs of particles over a plane, and then extending the analysis to the case where the triangular relation appears. This is qualitatively similar to the physics studied in \cite{Witten}. We will then analyze in detail the triangular relation which is based in the Quantum Yang Baxter equations (QYBE) \cite{Yang Baxterla}. In such scenario, the internal lines in the Yang-Baxter diagrams correspond to the pair of Goldstone bosons plus the degenerate vacuum. The external lines are labels corresponding to the $D$-branes in String Theory. Then the three spaces represented in the QYBE would correspond to the interaction of three coincident $D$-branes inside the framework of String Theory. This result then suggests that the Nambu-Goldstone theorem emerges naturally and fully from the scenario of String Theory, specifically from the interaction of gauge fields in Yang-Mills theories. The paper is organized as follows: In Sec. (\ref{eq:sec1}), we develop the basic concept of interaction of pairs of particles over a plane in some specific spacetime point. We then use these basic concepts for describing the way how pairs of Goldstone bosons might interact. In Sec. (\ref{s2}), we formulate the notion of spontaneous symmetry breaking as a triangular relation between pairs of Goldstone bosons and the degenerate vacuum. It will come out that this formulation is based on the QYBE if we introduce some auxiliary indices. In Sec. (\ref{s3}), we establish the direct connection between the triangular formulation of the spontaneous symmetry breaking and the QYBE. We then analyze the dispersion relations for the Goldstone bosons. In Sec. (\ref{s4}), we suggest the connection between the triangular version of the spontaneous symmetry breaking condition and the way how three coincident $D$-branes interact. The scenario suggest that there is a $1:1$ correspondence between the spaces in the QYBE and the $D$-branes. In addition, the quadratic dispersion relation for some Goldstone bosons comes out to be a consequence of the interaction of pairs of $D$-branes through the excitations represented by pairs of open Strings connecting the coincident normal coordinates. Finally, in Sec. (\ref{s5}), we conclude.                 

\section{The interaction of pairs of Nambu-Goldstone bosons}   \label{eq:sec1}

\begin{figure}
	\centering
		\includegraphics[width=17cm, height=8cm]{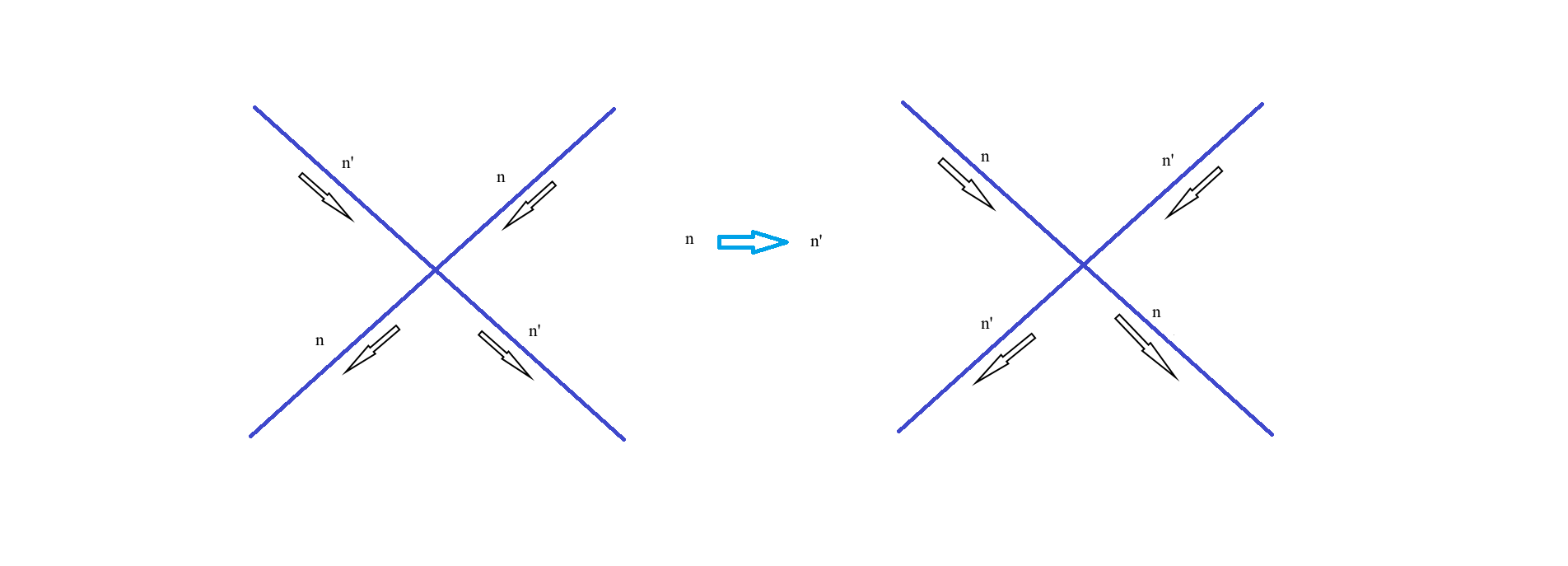}
	\caption{Pairs of Nambu-Goldstone bosons meeting at some location of spacetime. The slopes are related to the momentum of the particles.}
	\label{The titan3}
\end{figure}

If we consider the scenario where pairs of Goldstone bosons meet at some point in spacetime, then there is no loss of generality in working in two dimensions because any interaction of pairs of particles at a given spacetime location can be represented over a plane \cite{Witten}. Then for example, Fig. (\ref{The titan3}), illustrates the interaction of a pair of particles. The nature of the interaction is not relevant at this point. The slopes of the lines in the figure are related to the momentum of the particles. Then by keeping the slopes unchanged after the interaction, we are considering that there is no change in the momentum of the particles.   
The case illustrated on the left hand-side of Fig. (\ref{The titan3}), is the mirror image of the case illustrated on the right-hand side of the figure. Then we can say that both sides of the figure are twist map of each other. This can be observed from the fact that the right-hand side of the figure corresponds to the case where the roles of the particles are exchanged $n\to n'$. Both sides are the same up to some phases able to rotate the lines. For purposes of analysis, we will assume that the slopes of the particles $n$ and $n'$ are the same in magnitude but opposite in sign, then for the spatial momentum we will assume $\bf{p}_n=-\tilde{\bf{p}}_{n'}$. Then the exchange $n\to n'$ is clearly related to an inversion of the momentum of the particles. If the particles interacting are identical, then here we can also assume that $E_n=\tilde{E}_{n'}$. If we assume for example that the modes of the particles $n$ are related to the phase $e^{ip_nx}=e^{i(E_nx^0-\bf{p}_n\cdot\bf{x})}$ and that the modes of the particle $n'$ are related to the phase $e^{i\tilde{p}_{n'}x}=e^{i(\tilde{E}_{n'}x^0-\tilde{\bf{p}}_{n'}\cdot\bf{x})}$, then the exchange $n\to n'$ illustrated in Fig. (\ref{The titan3}), can be represented by the relations between phases 

\begin{equation}   \label{eq1}
e^{ip_nx}\to e^{-i(E_nx^0+\bf{p}_n\cdot\bf{x})}=e^{-i\tilde{p}_{n'}x},\;\;\;\;\;e^{i\tilde{p}_{n'}x}\to e^{-i(\tilde{E}_{n'}x^0+\tilde{\bf{p}}_{n'}\cdot\bf{x})}=e^{-ip_nx}.
\end{equation}  
Assuming that the particles just meet at the vertex without affecting their properties nor their motion, then the total event at the vertex would be a superposition of the histories illustrate in the figure. This issue will become relevant at the moment of understanding the principle of spontaneous symmetry breaking formulated as a triangular relation.    

\section{Spontaneous symmetry breaking as a triangular relation between pairs of Goldstone bosons and the vacuum}   \label{s2}

\begin{figure}
	\centering
		\includegraphics[width=20cm, height=8cm]{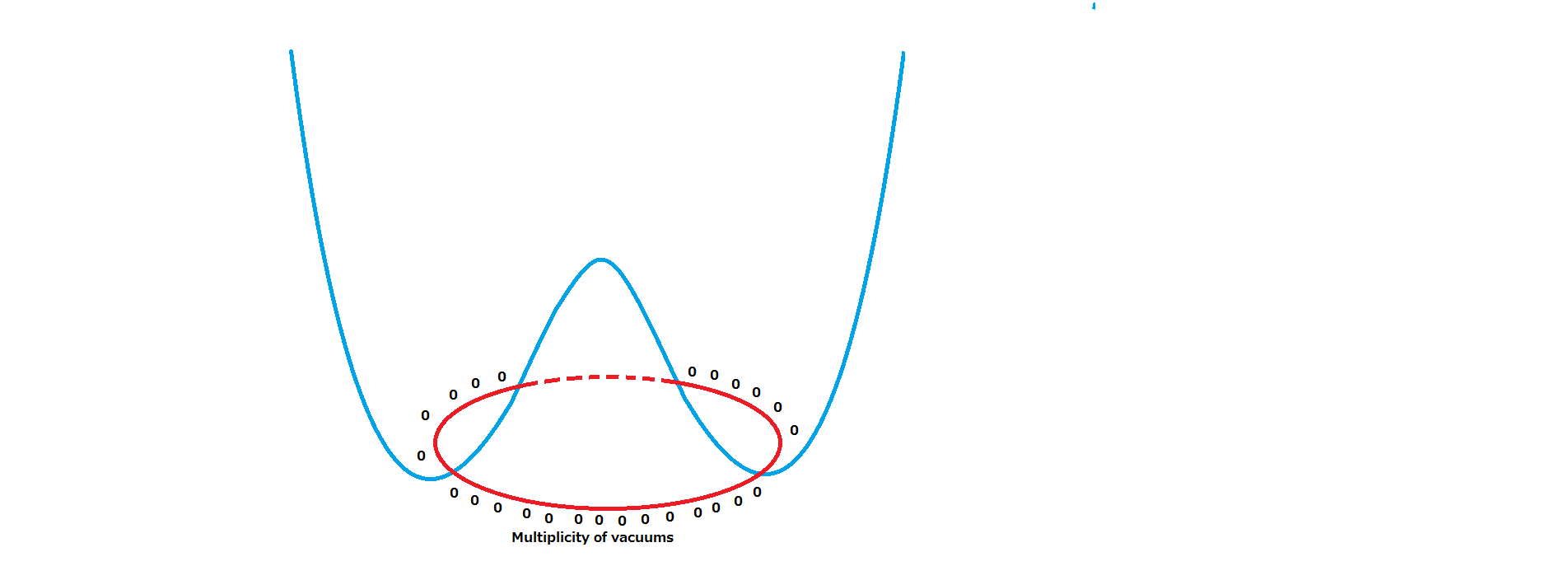}
	\caption{The vacuum degeneracy. In eq. (\ref{impo2}), we sum over the multiplicity of vacuums and then the expression related to the spontaneous symmetry breaking condition can vanish.}
	\label{The titan4}
\end{figure}

Given a system with two or more broken generators $Q_l(y)$, the spontaneous symmetry breaking condition can be defined as

\begin{equation}   \label{impo}
<0_{SV}\vert \left[\phi_{b}(x), [Q_{p}(y),Q_{l}(z)]\right]\vert0_{SV}>\neq0.
\end{equation} 
Here we are assuming that the pair of broken generators satisfy the Lie algebra, such that the previous expression resembles the standard form. We can sum eq. (\ref{impo}) over a degenerate vacuum, and then we obtain

\begin{equation}   \label{impo2}   
\sum_{0}<0_{DV}\vert \left[\phi_{b}(x), [Q_{p}(y),Q_{l}(z)]\right]\vert0_{DV}>=0.
\end{equation}   
Fig. (\ref{The titan4}) illustrates some possible situation where the symmetry is spontaneously broken. Note the presence of a multiplicity of vacuums (infinite vacuums) over which eq. (\ref{impo2}) is summed. The result (\ref{impo2}) can vanish without violating the spontaneous symmetry breaking condition \cite{YB}. The result (\ref{impo2}) illustrates the triangular relation between pairs of Goldstone bosons with the degenerate vacuum. In order to see this, we have to expand eq. (\ref{impo2}) in four terms defined, up to some phases as follows

\begin{eqnarray}   \label{impo3}
A=\sum_{0, n, n'}<0_{DV}\vert\phi_{b}(x)\vert n><n\vert\ Q_{p}(0)\vert n'><n'\vert Q_{l}(0)\vert0_{DV}>, \nonumber\\
B=\sum_{0, n, n'}<0_{DV}\vert\phi_{b}(x)\vert n'><n'\vert\ Q_{l}(0)\vert n><n\vert Q_{p}(0)\vert0_{DV}>, \nonumber\\
C=\sum_{0, n, n'}<0_{DV}\vert\ Q_{p}(0)\vert n><n\vert Q_{l}(0)\vert n'><n'\vert\phi_{b}(x)\vert 0_{DV}> \nonumber\\
D=\sum_{0, n, n'}<0_{DV}\vert Q_{l}(0)\vert n'><n'\vert\ Q_{p}(0)\vert n><n\vert\phi_{b}(x)\vert0_{DV}>.
\end{eqnarray}
Here we have introduced complete set of intermediate states defined as $\hat{I}=\vert n><n\vert=\vert n'><n'\vert$. Here $n$ and $n'$ makes reference to the pairs of Goldstone bosons. We have also assumed spacetime invariance for the broken generators as $Q_{p}(y)=e^{-ipy}Q_{p}(0)e^{ipy}$. However, in eq. (\ref{impo3}), we have not written explicitly the phases for each term. The phases will be fundamental in the coming analysis. Each term in eq. (\ref{impo3}) represents a triangular relation between pairs of Goldstone bosons ($n$ and $n'$) and the degenerate vacuum. The triangle relation is an extension of the interaction of pairs of particles illustrated in the previous section. Such relation is constrained by the QYBE defined in the same way as in the references \cite{YB, Witten}. We will demonstrate in the coming section that the triangle relations defined by the QYBE and illustrated in Fig. (\ref{Theking2}) are equivalent to

\begin{equation}   \label{impo4}
A=C,\;\;\;\;\;\;\;\;\;\;B=D.
\end{equation}
With this result, we can factorize each term in the commutator (\ref{impo2}), but taking into account the phases associated to the spacetime invariance. The commutator (\ref{impo2}), can be expanded as

\begin{eqnarray}   \label{AMI1}
\sum_{0}<0_{DV}\vert \left[\phi_{b}(x), [Q_{p}(y),Q_{l}(z)]\right]\vert0_{DV}>=A\left(e^{-i(p_n-p_{n'})y}e^{-i\tilde{p}_{n'}z}-e^{-i(\tilde{p}_{n}-\tilde{p}_{n'})z}e^{ip_{n}y}\right)\nonumber\\
+B\left(e^{-i(p_{n'}-p_{n})y}e^{i\tilde{p}_{n'}z}-e^{-i(\tilde{p}_{n'}-\tilde{p}_n)z}e^{-ip_{n}y}\right)=0.
\end{eqnarray} 
Later we will see how this result can help us to find the appropriate dispersion relation for the Goldstone bosons. 

\section{The triangular relation expressed as the QYBE}   \label{s3}

\begin{figure}
	\centering
		\includegraphics[width=17cm, height=15cm]{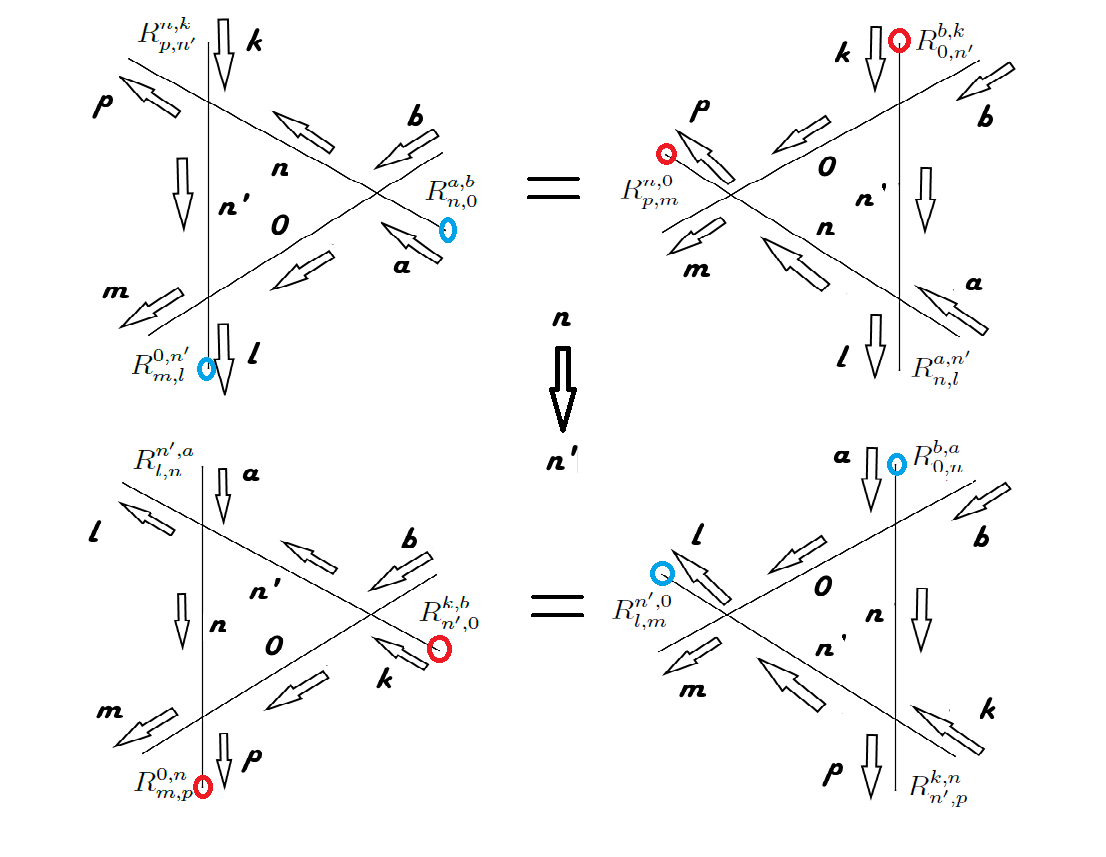}
	\caption{The effect of exchanging $n\to n'$. The upper relation, corresponds to the QYBE defined in eq. (\ref{eq:the last option2}). The lower relation corresponds to eq. (\ref{eq:the last optionmiau}). The figures marked with the same color are twist map (mirror image) of each other. The arrows indicate the flow of time. Taken from \cite{YB}.}
	\label{Theking2}
\end{figure}

The purpose of this section, is to understand the reason behind the factorization obtained in eq. (\ref{AMI1}). For that purpose, we have to introduce auxiliary indices in eqns. (\ref{impo2}) and (\ref{impo3}). The physical meaning of such auxiliary indices will become to be clear in the coming section where we explore the interactions of coincident $D$-branes inside the scenario of String Theory. It is possible to express the result described in eq. (\ref{impo4}) as a consequence of the standard QYBE. The condition $B=D$ for example, can be expressed in coordinate form as follows

\begin{equation}   \label{eq:the last option2}
R^{0,n'}_{m,l}R^{n,k}_{p,n'}R^{a,b}_{n,0}=R^{n,0}_{p,m}R^{a,n'}_{n,l}R^{b,k}_{0,n'},
\end{equation}
after introducing auxiliary indices. The $R$-matrices are defined as

\begin{eqnarray}   \label{another form3}
R^{0,n'}_{m,l}=<0_{DV}\vert Q_{m,l}(0)\vert n'>,\;\;\;R^{n,k}_{p,n'}=<n'\vert Q_{k,p}(0)\vert n>,\;\;\;\nonumber\\
R^{a,b}_{n,0}=<n\vert \phi_{a,b}(x)\vert0_{DV}>,\;\;\;R^{n,0}_{p,m}=<n\vert Q_{p, m}(0)\vert0_{DV}>,\;\;\;\nonumber\\
R^{a,n'}_{n,l}=<n'\vert Q_{l, a}(0) \vert n>,\;\;\;\;R^{b,k}_{0,n'}=<0_{DV}\vert \phi_{b, k}(x)\vert n'>.\;\;\;\;\;
\end{eqnarray}      
Under the exchange $n\to n'$, eq. (\ref{eq:the last option2}) becomes to be a different relation, defined as

\begin{equation}   \label{eq:the last optionmiau}
R^{0,n}_{m,p}R^{n',a}_{l,n}R^{k,b}_{n',0}=R^{n',0}_{l,m}R^{k,n}_{n',p}R^{b,a}_{0,n},
\end{equation}
which is equivalent to the result $A=C$ defined in eq. (\ref{impo4}) if we again introduce the corresponding auxiliary indices. Analogous definitions to the ones obtained in eq. (\ref{another form3}) appear for this case. Both relations, namely, the one illustrated in eq. (\ref{eq:the last option2}) together with the one illustrated in eq. (\ref{eq:the last optionmiau}) are showed graphically in Fig. (\ref{Theking2}). Note that the pair of Goldstone bosons as well as the degenerate vacuum become internal lines in the Yang-Baxter diagrams. This means that when we consider the spontaneous symmetry breaking condition, we have to take on equal footing the degenerate vacuum and the pairs of Goldstone bosons. 

\subsection{The dispersion relation of the Goldstone bosons}

In order to get the dispersion relation for the Goldstone bosons, we have to analyze how the momentum and frequency (energy) go to zero simultaneously. If the system distinguish between the exchange of the spacetime location for the Goldstone bosons, then $n\neq n'$ and we cannot factorize eq. (\ref{AMI1}) further. From the spatial integration over the space, together with the time-independence condition, the energy and momentum will approach simultaneously to zero linearly in eq. (\ref{AMI1}). Note from Fig. (\ref{Theking2}) that the exchange $n\to n'$ implies a change in the direction of flow over the phases (time reversal). In addition, the condition $n\neq n'$ is equivalent to say that the broken generators related to the pairs of Goldstone bosons $n$ and $n'$ are completely independent. The interesting case comes out to be the one where pairs of Goldstone bosons satisfy the condition $n=n'$. This is equivalent to say that the broken generators related to the pairs $n$ and $n'$, are canonically conjugate. In this case, from eqns. (\ref{impo3}), (\ref{impo4}), together with the related results (\ref{eq:the last option2}) and (\ref{eq:the last optionmiau}) we can conclude that 

\begin{equation}   \label{phases}
A=B=C=D.
\end{equation}    
This equality does not include the phases. We can notice that the arrows in the QYBE expressed graphically in Fig. (\ref{Theking2}), are advertising us that we have to make the appropriate changes in the sign of the phases when we factorize to a single expression the terms corresponding to the different QYBE. Note that we have two QYBE, defined as $A=C$ and $B=D$ in simplified notation (ignoring auxiliary indices). From the condition (\ref{phases}), we can simplify the result (\ref{AMI1}) as

\begin{equation}   \label{AMI2}
\sum_{0}<0_{DV}\vert \left[\phi_{b}(x), [Q_{p}(y),Q_{l}(z)]\right]\vert0_{DV}>=A\left([e^{-i\tilde{p}_{n'}z}+e^{i\tilde{p}_{n'}z}_{n\to n'}]-[e^{ip_{n}y}+e^{-ip_{n}y}_{n\to n'}]\right)=0,
\end{equation} 

where we have used the standard result $px=E_n x_0-\bf{p}_n\cdot\bf{x}$ and we have omitted the terms where differences of $4$-momentum appear, namely, those terms of the form $p_n-p_{n'}$ which will become equivalent to the unity. Those terms will only represent the equality $p_n=p_{n'}$ after spatial integration and after imposing the time-independent condition. The subindex $n\to n'$ reminds us that we have to make the appropriate changes in the phases in the corresponding terms in agreement with the convention explained in eq. (\ref{eq1}). Then we can express the result (\ref{AMI2}) as

\begin{eqnarray}
\sum_{0}<0_{DV}\vert [\phi_{b}(x), [Q_{p}(y),Q_{l}(z)]]\vert0_{DV}>=\nonumber\\
A\left(e^{-i\tilde{E}_{n'}z_0}\left(e^{-i\tilde{\bf{p}}_{n'}\cdot\bf{z}}+e^{i\tilde{\bf{p}}_{n'}\cdot\bf{z}}\right)-e^{iE_{n}y_0}\left(e^{i\bf{p}_n\cdot\bf{y}}+e^{-i\bf{p}_{n}\cdot\bf{y}}\right)\right)=0.
\end{eqnarray}     
This result can then be simplified into

\begin{equation}   \label{The equation la}
\sum_{0}<0_{DV}\vert [\phi_{b}(x), [Q_{p}(y),Q_{l}(z)]]\vert0_{DV}>=\nonumber\\
2A\left(e^{-i\tilde{E}_{n'}z_0}cos(\tilde{\bf{p}}_{n'}\cdot\bf{z})-e^{iE_n y_0}cos(\bf{p}_n\cdot\bf{y})\right)=0.
\end{equation}
We can then notice that the condition $n=n'$ implies that at the lowest order in the expansion, the energy (frequency) vanishes linearly and the spatial momentum vanishes quadratically. In addition, if $x_0=y_0$, as well as $\tilde{\bf{p}}_{n'}=-\bf{p}_n$, in agreement with the convention explained previously, then the previous result would vanish consistently in the limit $\bf{y}\to\bf{z}$. Note that the sign difference for the spatial momentum comes from the difference in the slopes between the lines corresponding to $n$ and $n'$ between the diagrams which are mirror image (twist map) of each other in Fig. (\ref{Theking2}). Compare for example the diagrams marked with blue circles in Fig. (\ref{Theking2}), as well as those marked with red circles. Compare the situation with the one illustrated in Fig. (\ref{The titan3}), where the vacuum lines are absent. Under the previous conditions, eq. (\ref{The equation la}) becomes

\begin{equation}   \label{The equation la2}
\sum_{0}<0_{DV}\vert [\phi_{b}(x), [Q_{p}(y),Q_{l}(z)]]\vert0_{DV}>=\nonumber\\
-4iAsin(E_n y_0)cos(\bf{p}_n\cdot\bf{y})=0,
\end{equation}
in the limit $E_n\to0$ and $\bf{p}_n\to0$. From this expression, the previous conclusions are more evident.        

\section{$D$-branes interactions and the Nambu-Goldstone theorem}   \label{s4}    
		
\begin{figure}
	\centering
		\includegraphics[width=25cm, height=8cm]{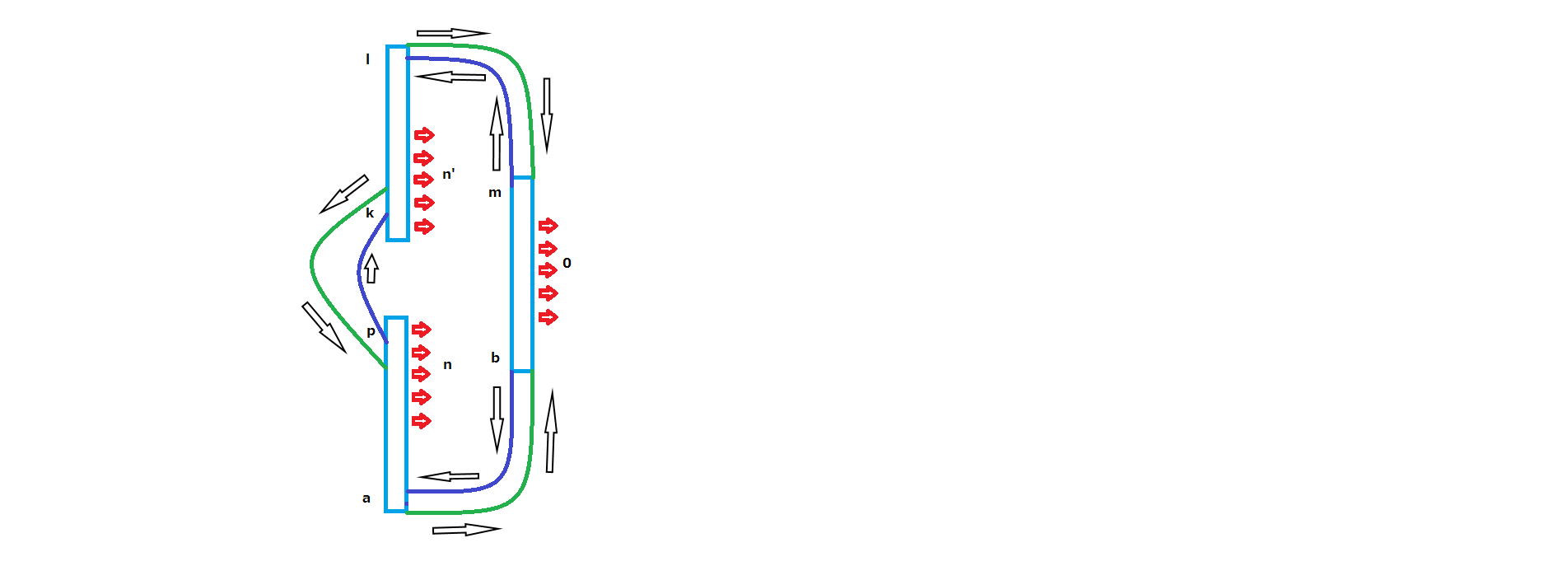}
	\caption{Schematic representation of six sectors for three coincident $D$-branes (Three trivial sectors omitted). For clarity, we show the $D$-branes separated from each other. Each $D$-brane would correspond to one space inside the QYBE. The arrows show the direction for each String sector. The pair of sectors joining the coincident normal coordinates originated from $p$ and $k$ are the reason for getting a quadratic dispersion relation for some Goldstone bosons.}
	\label{Theking21}
\end{figure}

At low energies, the results of the previous sections correspond to the interaction of $D$-branes in String Theory. Inside this scenario, the auxiliary indices assume the role of labels for the $D$-branes. Here we will not consider the Higgs mechanism since we are concerned about global symmetries. By considering the $D$-branes to be coincident, we avoid any possible dynamical origin for the mass of the gauge fields. The three spaces corresponding to each line in the QYBE illustrated in Fig. (\ref{Theking2}), would correspond in this scenario to three coincident $D$-branes. Then the group to be considered is the $U(3)$ Yang-Mills symmetry. This symmetry is closely related to the $SU(3)$ symmetry coming from the standard model Yang-Mills theory as follows \cite{Citeherela}.

\begin{equation}
U(3)=SU(3)\times U(1).
\end{equation}
Note that here $U(3)$ has nine sectors since in general the symmetry $U(N)$ has $N^2$ sectors. Here $SU(N)$ has $N^2-1$ sectors (eight for the present case) and finally $U(1)$ has one more sector, which is decoupled from the others. From Fig. (\ref{Theking21}), we can observe the scenario of three coincident $D$-branes interacting through the excitations represented by open Strings. In the figure the Strings starting and finishing in the same $D$-brane are omitted. Note that the normal coordinates for the pair of $D$-branes represented by the letters $l-k$ and $p-a$, represent the set of Goldstone bosons. The normal coordinates to the brane $m-b$ represent the multiplicity of vacuums, which we take on equal footing with the Goldstone bosons in this paper. The quadratic dispersion relation is a consequence of the superposition of the pair of open Strings starting and finishing in the same $D$-branes but flowing in the opposite direction between the pairs of $D$-branes $l-k$ and $p-a$. Such excitations join common normal coordinates for the pair of interacting $D$-branes. Note that the common normal coordinates for the pair of $D$-branes under consideration are represented with the letters $p$ and $k$ in the figure. A rigorous analysis for the scenario proposed in this paper will be developed in a subsequent work.     

\section{Conclusions}    \label{s5}

In this paper we have demonstrated the consistency of the triangular formulation based in the QYBE in order to formulate the spontaneous symmetry breaking condition. The triangle relation, constrained by the QYBE guarantees the appropriate dispersion relation as well as the appropriate number of Goldstone bosons. Although in this paper we did not mention the number of Goldstone bosons problem, it is evident from the previous analysis that under some circumstances, pairs of Goldstone bosons interact, becoming then effectively a single degree of freedom with quadratic dispersion relation. This is what happens when $n=n'$ in the QYBE, illustrated graphically in Fig. (\ref{Theking2}). Graphically the quadratic dispersion relation comes from the superposition of the graphics which are mirror images of each other. From the perspective of String Theory, we have three coincident $D$-branes, interacting through excitations represented by open Strings going from one $D$-brane to the other. The Goldstone bosons for some specific pairs of $D$-branes, would correspond to the excitations corresponding to normal coordinates of the $D$-brane. For coincident normal coordinates, represented by $p$ and $k$ in Fig. (\ref{Theking21}), there are pairs of excitations corresponding to open Strings starting in one $D$-brane and finishing in another. The pair of Strings have opposite orientation and their superposition gives us effectively a single excitation (Goldstone boson) with quadratic dispersion relation. This is the way how this result should be interpreted in String Theory. Note that three coincident $D$-branes in general represent nine sectors. In Fig. (\ref{Theking21}), we have omitted three sectors corresponding to the open Strings starting and finishing in the same $D$-brane. The figure only shows six sectors corresponding to the interactions of pairs of $D$-branes. From the six sectors, four correspond to the interaction of the Goldstone bosons with the vacuum. The remaining two, joining $p$ with $k$, correspond to the interactions of pairs of Goldstone bosons as has been explained before. Note the correspondence between the indices in the QYBE in Fig. (\ref{Theking2}), with the $D$-brane indices in Fig. (\ref{Theking21}). The result of this paper suggest that the spontaneous symmetry breaking condition, as well as the associated Nambu-Goldstone theorem, are natural consequences of String Theory, in particular, the Yang-Mills theories emerging from such scenario.     \\\\

{\bf Acknowledgement}
I thank the organizers of the conference "Quantum Gravity, String Theory and holography" organized in YITP, for the hospitality during my participation in the event, where a presentation about these ideas was done. I also thank Hitoshi Murayama for nice discussions about this idea, as well as the organizers of the conference "Why does the universe accelerates", for the support for attending this event organized at KEK, Tsukuba. I. A. is supported by the JSPS Post-Doctoral fellow for oversea researchers.

\end{document}